\documentstyle[12pt]{article}

\begin{document}
\newcommand{\bx}{\mbox{\boldmath $x$}}
\newcommand{\bF}{\mbox{\boldmath $F$}}
\newcommand{\bv}{\mbox{\boldmath $v$}}
\newcommand{\bz}{\mbox{\boldmath $z$}}
\newcommand{\bn}{\mbox{\boldmath $n$}}

\title{A Calculation on the Self-field of a Point Charge\\
and the Unruh Effect}

\author{Toru {\sc Hirayama}\thanks{E-mail: hira@cc.kyoto-su.ac.jp} 
and Tetsuya {\sc Hara}\thanks{E-mail: hara@cc.kyoto-su.ac.jp}\\
\small Department of Physics, Kyoto Sangyo University, Kyoto 603-8555, Japan}
\date{May, 2000}

\maketitle

\begin{abstract}
Within the context of quantum field theory in curved 
spacetimes, Hacyan and Sarmiento defined the vacuum stress-energy 
tensor with respect to the accelerated observer.  They calculated it 
for uniform acceleration and circular motion, and derived that 
the rotating observer perceives a flux.  Mane related the flux to 
synchrotron radiation.  In order to investigate the relation between 
the vacuum stress and bremsstrahlung, we estimate the stress-energy 
tensor of the electromagnetic field generated by a point charge, at the 
position of the charge.  We use the retarded field as 
a self-field of the point charge. Therefore the tensor diverges if we 
evaluate it as it is.  Hence we remove the divergent contributions 
by using the expansion of the tensor in powers of the distance 
from the point charge.  Finally, we take an average for the angular 
dependence of the expansion.  We calculate it for the case of
uniform acceleration and circular motion, and it is found that 
the order of the vacuum stress multiplied by $\pi\alpha$ 
($\alpha=e^2/\hbar c$ is the fine structure constant) is equal 
to that of the self-stress. In the Appendix, we give another 
trial approach with a similar result.
\end{abstract}

\newpage

\section{Introduction}
   
In Minkowski spacetime, a field is quantized with respect not 
only to the inertial frame, but also to a uniformly accelerated 
frame~\cite{rf:1}. Definitions of the vacua of these quantizations
are not equivalent.  In the quantization with respect to the uniformly
accelerated frame, 
the vacuum of the inertial frame corresponds to a thermal bath 
in which the temperature is proportional to the acceleration of
the accelerated frame.  Therefore, one can interpret that a uniformly
accelerated observer in the vacuum perceives a thermal bath of
temperature proportional to his acceleration~\cite{rf:2,rf:3}.  
This is referred to as the Unruh effect.   
This interpretation is confirmed by using the 
Unruh-DeWitt detector, which is a mathematically idealized 
detector of the field
quanta~\cite{rf:3,rf:4,rf:5,rf:6}.  It is well known that, 
when the detector is uniformly
accelerated, the transition probability between the internal states
of the detector indicate the thermal behavior.  The detector is also
excited in any accelerated motion, but, in general, the transition
probability does not indicate the thermal behavior.  The behavior of
the detector in a rotating orbit (circular Unruh effect~\cite{rf:7}) is
particularly interesting because of the possibility of experimental
verification~\cite{rf:8}.

It is interesting to conjecture how an accelerated electron can 
be affected by the
Unruh-like effect ascribed above. The spin of the electron could
correspond to the internal degree of freedom of the 
detector~\cite{rf:8},  and it is concerned with the 
experimental verificaton of the
circular Unruh effect.  On the other hand, it would be also 
interesting to investigate the relation
between bremsstrahlung and     
Unruh-like effect~\cite{rf:9,rf:10,rf:11,rf:12,rf:13},  and this 
is what we investigate in this paper.  
In this connection, there is a long-standing problem of 
classical electrodynamics concerning whether a uniformly accelerated 
electron radiates~\cite{rf:14,rf:15,rf:16},  
and it would be interesting
to consider this problem in connection with the Unruh effect.  
Discussions from this point of view are found, for example, 
in Refs.~\cite{rf:9,rf:12} and~\cite{rf:17}.

In this paper, we make a calculation 
within the classical theory~\cite{rf:18},
concerned with the discussion of Mane~\cite{rf:10}, which relates 
bremsstrahlung to the Unruh effect.  The outline of the discussion 
is the following.  Within quantum
field theory in curved spacetimes, Hacyan and Sarmiento defined the
spectrum of the stress-energy tensor of the electromagnetic
vacuum with respect to an accelerated observer, and calculated it for
uniformly accelerated motion~\cite{rf:19} and circular 
motion~\cite{rf:20}.  
For uniform acceleration, they obtained a spectrum of an isotropic
thermal bath, and for circular motion, they derived that the spectrum 
is not thermal, and there is a flux directed along the
tangent velocity of the observer.  Hacyan and Sarmiento pointed out
the possibility that the flux would cause some friction-like effect
on a rotating particle.  Mane suggested that this friction-like effect is
related to synchrotron radiation.  
Mane discussed that if the flux is coupled to an electron
through the fine structure constant $\alpha=e^2/\hbar c$, the order
of energy loss of the electron is classical, and it corresponds to the
order derived from the Larmor formula.

We are interested in how the
vacuum stress can be related to the classical bremsstrahlung, and we
propose to evaluate the stress-energy tensor of the electromagnetic field
generated by a point charge, at the position of the charge~\cite{rf:21}.
(We will call this quantity the self-stress only for simplicity,
although this term is generally used with a different meaning and context. 
See Section 17.5 of Ref.~\cite{rf:24}.)
We use the
retarded field as a self-field, and thus the tensor diverges if we evaluate
it as it is.  We consider the expansion of the tensor  
in powers of the distance from the point charge, and we
remove the divergent contributions in the limit that the distance
approaches zero.  That is, we regard the renormalized tensor as the
terms of zero-th order in the expansion.  Although the
result depends on the direction along which we take the limit, we remove
the directional dependence by taking an angular average. 
(In 1971, Teitelboim showed that the radiation reaction force of the
Lorentz-Dirac equation can be obtained by averaging the retarded field 
around a point charge~\cite{rf:22}.  Our method of averaging 
the retarded stress-energy tensor is the same as his method of 
averaging the retarded field.)  We calculate this average 
for uniform acceleration and circular motion, and it is 
found that the order of the vacuum stress multiplied by 
$\pi\alpha$ is equivalent to that of the self-stress which we calculate. 
In Appendix B, we give an alternative evaluation of the 
self-stress in which we use the expansion of the retarded field in powers
of the retarded time, and obtain a similar result.

In section \ref{vm}, we review the vacuum stress-energy tensor defined 
by Hacyan and Sarmiento and the discussion given by Mane.  
In section \ref{S3}, We expand the retarded field by using the method
of Dirac~\cite{rf;23} and briefly discuss the work of Teitelboim.
The expansion is used to evaluate 
the zero-th terms in the expansion of the stress-energy tensor of the
self-field in section \ref{S4}.  The result is disscused in section \ref{S5}.  
Throughout the paper, 
we use Gaussian units and the metric with signature $(+,-,-,-)$.  
We employ natural units in which $c=\hbar=1$, and we write $c$
and $\hbar$ explicitly only when an order estimation is needed.

\section{Vacuum stress and the discussion of Mane}
\label{vm}
\subsection{ Vacuum stress-energy tensor}

Within the context of quantum field theory in curved spacetimes, 
Hacyan and Sarmiento defined the electromagnetic vacuum stress-energy 
tensor with respect to the accelerated observer~\cite{rf:19,rf:20}.   
Let us review their work, mainly focusing on the points concerned with 
our problem.
The expectation value of the stress-energy tensor of the electromagnetic 
field is
\begin{eqnarray}
T_{\mu\nu}&=&\frac{1}{16\pi}\lim_{x'\rightarrow x}
\langle 0_M|
4F^\alpha_{(\mu}(x)F_{\nu)\alpha}(x')
+\eta_{\mu\nu}F_{\lambda\beta}(x)F^{\lambda\beta}(x')
|0_M\rangle,
\label{eq:1}
\end{eqnarray}
where $|0_M\rangle$ represents the Minkowski vacuum. 
We define 
\begin{eqnarray} 
D_{\mu\nu}^+(x,x')&\equiv&
\frac{1}{4}
\langle 0_M|
4F^\alpha_{(\mu}(x)F_{\nu)\alpha}(x')
+\eta_{\mu\nu}F_{\lambda\beta}(x)F^{\lambda\beta}(x')
|0_M\rangle,\nonumber\\
D_{\mu\nu}^-(x,x')&\equiv&D_{\mu\nu}^+(x',x),
\label{eq:2}
\end{eqnarray}
so that
\begin{eqnarray}
T_{\mu\nu}&=&
\frac{1}{4\pi}\lim_{x'\rightarrow x}D_{\mu\nu}^\pm(x,x'). 
\label{origin}
\end{eqnarray}
The decomposition of $F_{\mu\nu}$ into
the destruction and creation operators and the action of
these operators on $|0_M\rangle$ lead to the relation
\begin{eqnarray}
\langle 0_M|F^\alpha_{(\mu}(x)F_{\nu)\alpha}(x')|0_M\rangle
&=&
8\pi \partial_\mu \partial_{\nu'}D^\pm(x,x'),
\end{eqnarray}
where
\begin{eqnarray}
D^\pm(x,x')&=&
-\frac{1}{4\pi^2}\frac{1}{(t-t'\mp i\epsilon)^2-|\bx-\bx'|^2}
\end{eqnarray}
are the Wightman functions for the massless scalar field. By perfoming
the differentiations, we find that
\begin{eqnarray}
& &\langle 0_M|F^\alpha_{(\mu}(x)F_{\nu)\alpha}(x')|0_M\rangle
\nonumber\\
& &\hspace{2cm}=
\frac{4}{\pi}
\frac{4(x_\mu -x_\mu')(x_\nu -x_\nu' )
-\eta_{\mu\nu}(x_\alpha -x_\alpha')(x^\alpha -x'^\alpha )}
{[(t-t'\mp i\epsilon)^2-|\bx-\bx'|^2]^3}.
\label{eq:3} 
\end{eqnarray}
The contraction of the indices of this equation gives
\begin{eqnarray}
\langle 0_M|F_{\lambda\beta}(x)F^{\lambda\beta}(x')|0_M\rangle
&=&0,
\label{eq:4}
\end{eqnarray}
which leads to 
\begin{eqnarray}
T_{\mu\nu}&=&\frac{1}{4\pi}\lim_{x'\rightarrow x}
\langle 0_M|
F^\alpha_{(\mu}(x)F_{\nu)\alpha}(x')
|0_M\rangle.
\end{eqnarray}
That is, we find that the second term on the right-hand side of 
Eq.~(\ref{eq:1}) do not contribute to $T_{\mu\nu}$.

Now let us evaluate the spectrum of the stress-energy tensor detected 
by an observer through the world line
\begin{eqnarray}
z^\alpha&=&z^\alpha(\tau),
\end{eqnarray}
where $\tau$ is the proper time of the detector. It follows that
\begin{eqnarray}
T_{\mu\nu}[z^\alpha(\tau)]
&=&\frac{1}{4\pi}\int_{-\infty}^{\infty}d\sigma\delta(\sigma)
D_{\mu\nu}^+(\tau+\sigma/2,\tau-\sigma/2)\nonumber\\
& &
\hspace{-1.2cm}
=\frac{1}{8\pi^2}
\int_{-\infty}^{\infty}d\sigma\int_0^{\infty}d\omega
e^{i\omega\sigma}
[D_{\mu\nu}^+(\tau+\sigma/2,\tau-\sigma/2)
+D_{\mu\nu}^-(\tau+\sigma/2,\tau-\sigma/2)],
\nonumber\\
\label{totyu}
\end{eqnarray}
where
\begin{eqnarray}
D_{\mu\nu}^\pm(\tau+\sigma/2,\tau-\sigma/2)&\equiv&
D_{\mu\nu}^\pm(z(\tau+\sigma/2),z(\tau-\sigma/2)). 
\end{eqnarray}
Using the Fourier transform
\begin{eqnarray}
\tilde{D}_{\mu\nu}^\pm(\tau,\omega)&=&
\int_{-\infty}^{\infty}d\sigma
e^{i\omega\sigma}D_{\mu\nu}^\pm(\tau+\sigma/2,\tau-\sigma/2),
\end{eqnarray}
the stress-energy tensor can be written as
\begin{eqnarray}
T_{\mu\nu}[z^\alpha(\tau)]&=&
\frac{1}{8\pi^2}\int_0^{\infty}
[\tilde{D}_{\mu\nu}^{+}(\tau,\omega) 
+ \tilde{D}_{\mu\nu}^{-}(\tau,\omega)]
d\omega.\label{eq:6}
\end{eqnarray}
  
Because the $D_{\mu\nu}^\pm(\tau+\sigma/2,\tau-\sigma/2)$ are even 
functions with respect to $\sigma$, and because of Eqs.~(\ref{eq:2}),
(\ref{eq:3}) and~(\ref{eq:4}),
we can write 
\begin{eqnarray}
D_{\mu\nu}^\pm(\tau+\sigma/2,\tau-\sigma/2)&=&
\frac{A_{\mu\nu}}{(\sigma\mp i\epsilon)^4}
-\frac{B_{\mu\nu}}{(\sigma\mp i\epsilon)^2}
+D_{\mu\nu}(\tau,\sigma),\label{eq:5}
\end{eqnarray}
where $A_{\mu\nu}$ and $B_{\mu\nu}$ are functions of $\tau$, and
$D_{\mu\nu}(\tau,\sigma)$ is by definition free of poles at 
$\sigma=\pm i\epsilon$. Inserting Eq.~(\ref{eq:5}) into 
Eq.~(\ref{eq:6}), we obtain that
\begin{eqnarray}
T_{\mu\nu}[z^\alpha(\tau)]&=&
\frac{1}{8\pi^2}\int_0^{\infty}
\left[
\frac{A_{\mu\nu}}{3}\omega^3+2B_{\mu\nu}\omega
\right]d\omega
+\frac{1}{4\pi}D_{\mu\nu}(\tau,0).\label{eq:7}
\end{eqnarray}
One can interpret this equation is expressing that 
the divergent integral term corresponds to
the zero-point energy, and the last term gives the physically 
observable stress-energy tensor.

We can use Eq.~(\ref{eq:6}) to obtain the spectrum of the stress-energy 
tensor. For example, 
for uniform acceleration with acceleration $a$, 
Eq.~(\ref{eq:6}) turns out to be
\begin{eqnarray}
T_{\mu\nu}&=&
\frac{1}{3\pi^2}(4u_\mu u_\nu-\eta_{\mu\nu})
\int_{0}^{\infty}\omega(\omega^2+a^2)
\left[\frac{1}{2}+\frac{1}{e^{2\pi\omega/a}-1}\right]d\omega,
\label{Planck}
\end{eqnarray}
where $u_\mu$ is the 4-velocity of the observer. 
If one considers that the effect of the acceleration changes the density of 
states from
$\omega^2d\omega$ to $(\omega^2+a^2)d\omega$ [This change is 
clarified by the spin of the field. see Ref.~\cite{rf:19}.], the above 
spectrum can be 
interpreted as a Planck spectrum. The term $\omega(\omega^2+a^2)/2$ is 
considered as the zero-point energy of the field in the accelerated frame,
and this just corresponds to the divergent term of 
Eq.~(\ref{eq:7}).

Hacyan and Sarmiento pointed out that one can remove the divergent
contribution of the vacuum stress-energy tensor by moving the pole of 
the Wightman function properly. But they 
judged that this is a rather {\it ad hoc} procedure, and they 
decided not to discard the zero-point energy in Eq.~(\ref{eq:7})
(see section V of Ref.~\cite{rf:20}).
Thus they were careful with divergence elimination. 
However, they regarded virtually $(4\pi)^{-1}D_{\mu\nu}(\tau,0)$
as the renormalized stress-energy tensor. Therefore we also adopt
it as the renormalized vacuum stress-energy tensor.

Here we give a simple interpretation of their renormalization 
procedure to contrast it with our renormalization procedure of
the classical self-stress, which will be introduced later. 
The interpretation is as follows.
One first expands $T_{\mu\nu}$ in powers of the proper time as in 
Eq.~(\ref{eq:5}), and then removes contributions which diverge when 
$\sigma \rightarrow 0$. Doing so, one obtains the contribution with 
a zero-th order as the result.

Finally, let us note the results for the renormalized vacuum stress
for uniform acceleration and circular motion 
(Eqs.~(3.6) and~(4.23) in Ref.~\cite{rf:20}). 
The world line of the 
uniformly accelerated observer is
\begin{eqnarray}
z^\mu&=&
(a^{-1}\sinh (a\tau),0,0,a^{-1}\cosh (a\tau)).
\label{eq:14}
\end{eqnarray}
If one evaluates the vacuum stress at the instant that the observer 
is at rest, i.e., at $\tau=0$, the result is 
\begin{eqnarray}
T^{\mu\nu}=\frac{11}{720\pi^2}\frac{\hbar a^4}{c^7}
\left[
\begin{array}{cccc}
3&0&0&0\\
0&1&0&0\\
0&0&1&0\\
0&0&0&1
\end{array}
\right].
\label{Quniform}
\end{eqnarray}

In the laboratory frame, circular motion can be written 
\begin{eqnarray}
z^\mu&=&
(\gamma\tau, R\cos(\gamma\Omega\tau),R\sin(\gamma\Omega\tau),0),
\label{circ}
\end{eqnarray}
where $R$ is the radius of the circle, $v$ is the velocity in the 
laboratory frame, $\Omega=v/R$, and $\gamma=(1-v^2)^{-1/2}$.
Here we define 
\begin{eqnarray}
k^\mu&=&
(1,0,0,0),\nonumber
\\
\tilde{l}_1^\mu&=&
(0, \cos(\gamma\Omega\tau),\sin(\gamma\Omega\tau),0),\nonumber
\\ 
\tilde{l}_2^\mu&=&
(0,-\sin(\gamma\Omega\tau),\cos(\gamma\Omega\tau),0),\nonumber
\\
\tilde{l}_3^\mu&=&
(0,0,0,1).
\end{eqnarray}
Components of these vectors are defined in the laboratory frame.
They are orthonormal. We can write 
\begin{eqnarray}
\dot{z}^\mu
&=&
\gamma k^\mu+\gamma v \tilde{l}_2^\mu.
\end{eqnarray}
Then, we also define 
\begin{eqnarray}
l_1^\mu&=&\tilde{l}_1^\mu,\nonumber
\\ 
l_2^\mu&=&
\gamma v k^\mu+\gamma\tilde{l}_2^\mu,\nonumber
\\
l_3^\mu&=&\tilde{l}_3^\mu.
\end{eqnarray}
The Lorentz transformations of $k^\mu$, $\tilde{l}_1^\mu$, 
$\tilde{l}_2^\mu$ and $\tilde{l}_3^\mu$ with respect to the velocity 
of the observer are
$\dot{z}^\mu$, $l_1^\mu$, $l_2^\mu$ and $l_3^\mu$. Therefore 
$\dot{z}^\mu$, $l_1^\mu$, $l_2^\mu$ and $l_3^\mu$ are orthonormal, and
they constitute the coordinate basis of the rest frame of the observer. 
We set the
 $x$, $y$ and $z$ axes along the directions of $l_1^\mu$, $l_2^\mu$ and  
$l_3^\mu$.
The vacuum stress is evaluated, 
in the rest frame of the observer, as
\begin{eqnarray}
T^{\mu\nu}&=&
\frac{1}{1440\pi^2}\frac{\hbar\gamma^8\Omega^4v^2}{c^5}
\nonumber\\
& &\times
\left[
\begin{array}{cccc}
100-66\gamma^{-2}&0&(50-47\gamma^{-2})c/v&0\\
0&30-22\gamma^{-2}&0&0\\
(50-47\gamma^{-2})c/v&0&40-22\gamma^{-2}&0\\
0&0&0&30-22\gamma^{-2}
\end{array}
\right].
\nonumber\\
\label{eq:8}
\end{eqnarray}
(It seems that Eqs.~(4.23a-c) in Ref.~\cite{rf:20} are misprinted.) 
We should note that the Poynting vector is not zero, and the flux 
is directed along the $y$ axis, i.e., along the Lorentz boost from the 
laboratory frame to the rest frame of the observer. Hacyan and 
Sarmiento pointed out that ``if this flux is real, it should imply 
some friction-like effect on a rotating particle''.

\subsection{{\it The discussion of Mane}}
 
Mane suggested that the flux is related to the synchrotron 
radiation~\cite{rf:10}. We outline his discussion here.
We consider the case where a charged particle is moving along the 
orbit with radius of curvature $R$ in the ultrarelativistic limit.

First we consider {\it the area in which the charged particle interacts 
with the electromagnetic field} 
(see section 14.4 of Ref.~\cite{rf:24}).  Because an 
ultrarelativistic particle radiates with angle 
$\theta\sim\gamma^{-1}$, the observer at rest at infinity 
observes the radiation mainly during the time that the particle 
rotates by 
$\theta\sim\gamma^{-1}$. Therefore the particle interacts with the 
electromagnetic field in the time 
$\Delta t \sim (R\theta)/v \sim R/(\gamma v)$. 
In that time, the radiation travels a distance 
$D = c\Delta t \sim (Rc)/(\gamma v)$. Therefore the electromagnetic 
wave radiated in $\Delta t$ spreads out in the area 
$A \sim (\theta D)^2$ on a surface orthogonal to the orbit. 
This area is invariant for the Lorentz transformation from the 
laboratory frame to the rest frame, because the surface is orthogonal 
to the orbit. We regard $A$ as {\it the area in which the charged 
particle interacts with the electromagnetic field}.

We can write the Poynting flux in Eq.~(\ref{eq:8})
in the form
\begin{eqnarray}
pl_2^\mu&=&
\frac{1}{1440\pi^2}
\frac{\hbar \gamma^8\Omega^4 v}{c^4}(50-47\gamma^{-2})l_2^\mu,
\end{eqnarray}
where $l_2^\mu$ is given as
\begin{eqnarray}
l_2^\mu&=&
((\gamma v)/c,-\gamma\sin (\gamma\Omega\tau),
\gamma\cos (\gamma\Omega\tau),0).
\end{eqnarray}
Since $p$ is proportional to $\hbar$, 
the flux becomes zero in the classical limit.
But if the particle couples with the flux by the fine structure 
constant $\alpha = e^2/\hbar c$, the contribution of $\hbar$ vanishes, 
and the effect on the particle becomes classical. The recoil 
induced by the flux of the vacuum fluctuation 
on the four-momentum of the particle per unit proper time is
\begin{eqnarray} 
\alpha A pl_2^\mu
&\sim& 
\frac{e^2\gamma^4\Omega^2 v}{c^3}l_2^\mu.
\end{eqnarray}

In the laboratory frame, the energy loss of the particle per unit 
laboratory time is given by the Larmor formula
\begin{eqnarray} 
I&=&\frac{2}{3}\frac{e^2}{c^3}(\gamma^2\Omega v)^2.
\end{eqnarray}
This is related to the damping force $\bF$ in the form 
$I = \bF\cdot\bv$ and therefore the recoil induced by 
synchrotron radiation
on the four-momentum of the particle per unit proper time is
\begin{eqnarray} 
\gamma (I/c,\bF)
&\sim&
\frac{e^2\gamma^4\Omega^2 v}{c^3}l_2^\mu.
\end{eqnarray}
Hence, we find that, 
if one assume that the charged particle interacts with the vacuum flux 
by the coupling $\alpha$, the order of recoil of the particle 
induced by this interaction is equal to that 
derived by the Larmor formula in the ultrarelativistic limit.

\section{Expansion of a retarded field}
\label{S3}

To investigate the relation between the vacuum stress and the 
bremsstrahlung,
we evaluate the stress-energy tensor of a self-field generated by 
a charged particle, at the position of the particle. 
We use a point charge and adopt the retarded field as the self-field. 
Therefore, the self-stress is now divergent if we evaluate it as it is. 
Hence we must remove the divergent contributions with some procedure. 
First, we construct the expansion of the retarded field in powers of 
the distance from the point charge. By doing this, we calculate the 
terms of zero-th order in the expansion of the stress-energy tensor, 
and we regard the result as the renormalized stress-energy tensor.
This procedure reminds us of our interpretation of the renormalization 
of the vacuum stress, where we expanded the stress-energy tensor
in powers of the proper time. 

In calculating the expansion of the field, we adopt the method used 
by Dirac in Ref.~\cite{rf;23}, which is followed in this section 
and in Appendix A. (Dirac used the expansion for 
calculating the energy-momentum
flow out of the world tube surrounding the world line of a point charge. 
However, our aim is not to investigate this quantity. Our aim is to 
evaluate the each component of $T_{\mu\nu}$.)

The retarded potential generated by a 4-current $j^\mu(x)$ is given 
in the form
\begin{eqnarray}
A^\mu_{{\rm ret}} (x)&=&
4\pi\int d^4x'D_r(x-x')j^\mu(x'),
\label{eq:11}
\end{eqnarray}
where 
\begin{eqnarray}
D_r(x-x')&=& 
\frac{1}{2\pi}\theta(x_0-x'_0)\delta[(x-x')^2]
\label{eq:9}
\end{eqnarray}
is the retarded Green function.
We define the world line of the point charge as $z^\mu(\tau)$, 
where $\tau$ is the proper time. Then the 4-current of the point 
charge with charge $e$ is
\begin{eqnarray}
j^\mu(x)&=&e\int_{-\infty}^{\infty}
d\tau\dot{z}^\mu(\tau)\delta^{(4)}[x-z(\tau)].
\label{eq:10}
\end{eqnarray}
The dot above $z$ represents differentiation with respect to the 
proper time.
Substituting Eqs.~(\ref{eq:9}) and~(\ref{eq:10}) into  
Eq.~(\ref{eq:11}), we have
\begin{eqnarray}
A^\mu_{{\rm ret}}(x)&=&
2e\int_{-\infty}^{\infty}
d\tau\dot{z}^\mu\theta(x_0-z_0)\delta[(x-z)^2]\nonumber\\
&=&2e\int d\tau\dot{z}^\mu\delta[(x-z)^2],
\end{eqnarray}
where, in the last, integration is taken 
from $-\infty$ to some value of $\tau$ intermadiate between 
the retarded and advanced times.
We now have
\begin{eqnarray}
\partial_\nu A_{\mu,{\rm ret}}(x)&=&
4e\int d\tau\dot{z}_\mu (x_\nu-z_\nu)\delta'[(x-z)^2]\nonumber\\
&=&
-2e\int d\tau\frac{\dot{z}_\mu (x_\nu-z_\nu)}{\dot{z}\cdot(x-z)}
\frac{d}{d\tau}\delta[(x-z)^2]\nonumber\\
&=&
2e\int d\tau\frac{d}{d\tau}
\left[\frac{\dot{z}_\mu (x_\nu-z_\nu)}{\dot{z}\cdot(x-z)}\right]
\delta[(x-z)^2].
\end{eqnarray}
Thus the retarded field of the point charge becomes
\begin{eqnarray}
F_{\mu\nu,{\rm ret}}(x)&=&
\partial_\mu A_{\nu,{\rm ret}}(x)-\partial_\nu A_{\mu,{\rm ret}}(x)
\nonumber\\
&=&
-2e\int d\tau\frac{d}{d\tau}
\left[
\frac{\dot{z}_\mu (x_\nu-z_\nu)-\dot{z}_\nu (x_\mu-z_\mu)}{\dot{z}\cdot(x-z)}
\right]
\delta[(x-z)^2]\nonumber\\
&=&
\frac{e}{\dot{z}\cdot(z-x)}\frac{d}{d\tau}
\frac{\dot{z}_\mu (z_\nu-x_\nu)-\dot{z}_\nu (z_\mu-x_\mu)}{\dot{z}\cdot(z-x)},
\label{eq:12}
\end{eqnarray}
where $z^\mu$ is evaluated at the retarded time 
in the last equation.

Here we set
\begin{eqnarray}
x^\mu&=&z^\mu(\tau_0)+\gamma^\mu
\end{eqnarray}
and expand Eq.~(\ref{eq:12}) in powers of $\gamma^\mu$. 
At that time, we choose
$\tau_0$ to satisfy
\begin{eqnarray}
\dot{z}(\tau_0)\cdot\gamma&=&0.
\end{eqnarray}
If one choose the frame in which the charge is instantaneously at rest at 
the instant $\tau=\tau_0$, $\gamma^{\mu}$ has only spatial components, 
so that $x^{0} = z(\tau_0)^{0}$. Also,
$\epsilon\equiv\sqrt{-\gamma\cdot\gamma}$ is the distance
 from $\bz(\tau_0)$ to 
$\bx$ in this frame. Therefore, the expansion with 
respect to  $\epsilon$ is equivalent to that with respect 
to the distance from the point charge in the instantaneous 
rest frame of the point charge. We point out that Dirac calculated 
the expansion of $[1-\gamma\cdot\ddot{z}]^{1/2}F_{\mu\nu,{\rm ret}}$
to obtain the energy-momentum flow out of the world tube, 
but we expand $F_{\mu\nu,{\rm ret}}$, because our purpose is
different from that of Dirac. The details of the culculation are 
complicated, and therefore they are given in Appendix A. 
Before we give the result of the calculation, some 
notation is defined:
\begin{eqnarray}
n^\mu&=&\epsilon^{-1}\gamma^\mu,\nonumber\\
(m)^\mu&=&\frac{d^mz^\mu}{d\tau^m},\nonumber\\
\Delta_m&=&n\cdot(m),\nonumber\\
\alpha_2&=&(2)\cdot(2),\nonumber\\
\alpha_3&=&(3)\cdot(3).
\end{eqnarray}

The expansion of $F_{\mu\nu,{\rm ret}}$ in powers of $\epsilon$
is derived as follows:
\begin{eqnarray}
F_{\mu\nu,{\rm ret}}(x)&=&
ef^{(-2)}_{\mu\nu}\epsilon^{-2}+
ef^{(-1)}_{\mu\nu}\epsilon^{-1}+
ef^{(0)}_{\mu\nu}+
ef^{(1)}_{\mu\nu}\epsilon+
ef^{(2)}_{\mu\nu}\epsilon^{2}
-(\mu\leftrightarrow\nu)
+O(\epsilon^3),\nonumber\\
f^{(-2)}_{\mu\nu}
&=&
n_\mu(1)_\nu,
\nonumber\\
f^{(-1)}_{\mu\nu}
&=&
\frac{1}{2}\Delta_2n_\mu(1)_\nu-\frac{1}{2}(2)_\mu(1)_\nu, 
\nonumber\\
f^{(0)}_{\mu\nu}
&=&
\left[
\frac{3}{8}(\Delta_2)^2-\frac{1}{8}\alpha_2
\right]n_\mu(1)_\nu
-\frac{3}{4}\Delta_2(2)_\mu(1)_\nu
-\frac{1}{2}n_\mu(3)_\nu
+\frac{2}{3}(3)_\mu(1)_\nu,
\nonumber\\
f^{(1)}_{\mu\nu}
&=&
\left[
\frac{5}{16}(\Delta_2)^3-\frac{5}{16}\Delta_2\alpha_2
-\frac{1}{8}\Delta_4+\frac{1}{6}\dot{\alpha}_2
\right]n_\mu(1)_\nu
+
\left[
-\frac{1}{2}\Delta_3+\frac{1}{3}\alpha_2
\right]n_\mu(2)_\nu\nonumber\\
& &+
\left[
-\frac{15}{16}(\Delta_2)^2+\frac{2}{3}\Delta_3-\frac{5}{16}\alpha_2
\right](2)_\mu(1)_\nu
-\frac{3}{4}\Delta_2n_\mu(3)_\nu
\nonumber\\
& &+\frac{4}{3}\Delta_2(3)_\mu(1)_\nu+\frac{1}{3}n_\mu(4)_\nu
-\frac{3}{8}(4)_\mu(1)_\nu-\frac{1}{4}(3)_\mu(2)_\nu,
\nonumber\\
f^{(2)}_{\mu\nu}
&=&
\left[
\frac{35}{128}(\Delta_2)^4-\frac{35}{64}(\Delta_2)^2\alpha_2
-\frac{5}{16}\Delta_2\Delta_4+\frac{1}{2}\Delta_2\dot{\alpha}_2
-\frac{5}{24}(\Delta_3)^2-\frac{35}{384}(\alpha_2)^2
\right.
\nonumber\\
& &\left.
+\frac{1}{3}\Delta_3\alpha_2+\frac{1}{15}\Delta_5
-\frac{3}{32}\ddot{\alpha}_2+\frac{1}{48}\alpha_3
\right]n_\mu(1)_\nu
\nonumber\\
& &+
\left[
-\frac{5}{4}\Delta_2\Delta_3+\Delta_2\alpha_2+\frac{1}{3}\Delta_4
-\frac{5}{16}\dot{\alpha}_2
\right]n_\mu(2)_\nu
\nonumber\\
& &+
\left[
-\frac{35}{32}(\Delta_2)^3+2\Delta_2\Delta_3+\frac{1}{4}\dot{\alpha}_2
-\frac{5}{16}\Delta_4-\frac{35}{32}\Delta_2\alpha_2
\right](2)_\mu(1)_\nu
\nonumber\\
& &+
\left[
-\frac{15}{16}(\Delta_2)^2+\frac{2}{3}\Delta_3
-\frac{5}{16}\alpha_2
\right]n_\mu(3)_\nu
+
\left[
2(\Delta_2)^2-\frac{5}{6}\Delta_3+\frac{1}{3}\alpha_2
\right](3)_\mu(1)_\nu
\nonumber\\
& &+
\frac{2}{3}\Delta_2n_\mu(4)_\nu
-\frac{15}{16}\Delta_2(4)_\mu(1)_\nu
-\frac{5}{8}\Delta_2(3)_\mu(2)_\nu
\nonumber\\
& &
-\frac{1}{8}n_\mu(5)_\nu
+\frac{2}{15}(5)_\mu(1)_\nu+\frac{1}{6}(4)_\mu(2)_\nu
\label{eq:13},
\end{eqnarray}
where the functions of $\tau$ in the expansion are evaluated 
at the time $\tau=\tau_0$.

Here we would like to point out the work of Teitelboim~\cite{rf:22}, 
who showed that
the radiation reaction force of the Lorentz-Dirac equation can be derived 
by simply averaging the angular dependence of above expansion of the field.
We summarize his discussion in the following.

He proposed to evaluate the value of the retarded field at the 
particle's own position and the force acting on the particle.
However, obviously there are two problems. The first is that the 
retarded field 
diverges at the position of the particle. The second is that the ``limit''
of the retarded field depends on the direction along which the 
singularity is approached. In fact, in
Eq.~(\ref{eq:13}), the angular dependence, $n^\mu$, is included in the 
coefficients of the expansion. 

He avoided the second problem by simply 
averaging the angular dependence of Eq.~(\ref{eq:13}) in the 
instantaneous rest frame of the charge. In this frame, one can write 
$n^\mu = (0,n_x,n_y,n_z)$, so that $\bn=(n_x,n_y,n_z)$ is the unit 
vector directed from the position of the charge to reference point of 
the field. He averaged Eq.~(\ref{eq:13}) to the order of 
$O(\epsilon^0)$. The terms in Eq.~(\ref{eq:13}) 
which contain odd $n^\mu$s vanish when the average is performed, because 
the signs of these terms change when the direction of $\bn$ is 
reversed.  Only remained term is 
$\Delta_2n_\mu (1)_\nu$ in $f^{(-2)}_{\mu\nu}$. If one expresses an angular
averaged fuction by drawing a bar over the quantity, it follows that
\begin{eqnarray}
\overline{\Delta_2\bn}&=&
-\overline{(\bn\cdot\ddot{\bz})\bn}=
-\overline{\cos^2\theta}\ddot{\bz}=
-\frac{1}{3}\ddot{\bz},
\end{eqnarray}
where $\theta$ is the angle between $\ddot{\bz}$ and $\bn$. 
This relation is easily rewritten in the covariant form
\begin{eqnarray}
\overline{\Delta_2 n^\mu}&=&-\frac{1}{3}(2)^\mu.
\end{eqnarray}
Then it follows that
\begin{eqnarray}
\overline{F_{\mu\nu,{\rm ret}}}&=&
-\frac{2e}{3}(2)_\mu (1)_\nu\epsilon^{-1}+
\frac{2e}{3}(3)_\mu (1)_\nu-(\mu\leftrightarrow\nu)+O(\epsilon).
\label{bekibeki}
\end{eqnarray}
Thus the Lorentz force acting on the charge is obtained as
\begin{eqnarray}
e\overline{F_{\rm ret}^{\mu\nu}}(1)_\nu
&=&
-\left(
\lim_{\epsilon\rightarrow 0}\frac{2e^2}{3\epsilon}(2)^\mu
\right)
+\frac{2e^2}{3}\{(3)^\mu+\alpha_2(1)^\mu\}.
\end{eqnarray}
The first term on the right-hand side, which diverges in the limit 
$\epsilon\rightarrow 0$, is interpreted as the infinite Coulomb mass of 
the point charge, and this is absorbed in the usual way into the 
observed finite mass of the particle. The second term represents
the radiation reaction force, which is equivalent to the radiation 
reaction force of the Lorentz-Dirac equation~\cite{rf:22}.

\section{Evaluation of self-stress}
\label{S4}

In this section we calculate the terms with zero-th order in the 
expansion of the stress-energy tensor for a point charge with  
uniform acceleration and circular motion. We evaluate them in 
the rest frame of the charge at the instant $\tau=\tau_0$.

First let us calculate for the case of a point charge with uniform 
acceleration $a$. The world line of the charge is expressed 
by Eq.~(\ref{eq:14}). The charge is at rest at $\tau=0$, so we 
evaluate it at $\tau=0$. Inserting
\begin{eqnarray}
(2)^\mu&=&(0,0,0,a)\equiv am^{\mu},\nonumber\\
(3)^\mu&=&a^2(1)^\mu,\nonumber\\
(4)^\mu&=&a^3m^\mu,\nonumber\\
(5)^\mu&=&a^4(1)^\mu
\end{eqnarray}
into Eq.~(\ref{eq:13}), we have
\begin{eqnarray}
F_{\mu\nu,{\rm ret}}&=&
ef^{(-2)}_{\mu\nu,{\rm uni}}\epsilon^{-2}+
ef^{(-1)}_{\mu\nu,{\rm uni}}a\epsilon^{-1}+
ef^{(0)}_{\mu\nu,{\rm uni}}a^2+
ef^{(1)}_{\mu\nu,{\rm uni}}a^3\epsilon+
ef^{(2)}_{\mu\nu,{\rm uni}}a^4\epsilon^2\nonumber\\
& &-(\mu\leftrightarrow\nu)+O(\epsilon^3),\nonumber\\
f^{(-2)}_{\mu\nu,{\rm uni}}&=&
n_\mu(1)_\nu,
\nonumber\\
f^{(-1)}_{\mu\nu,{\rm uni}}
&=&\frac{1}{2}(n\cdot m)n_\mu(1)_\nu-\frac{1}{2}m_\mu(1)_\nu,
\nonumber\\
f^{(0)}_{\mu\nu,{\rm uni}}
&=&
\frac{3}{8}\left[-1+(n\cdot m)^2\right]n_\mu(1)_\nu
-
\frac{3}{4}(n\cdot m)m_\mu(1)_\nu,
\nonumber\\
f^{(1)}_{\mu\nu,{\rm uni}}
&=&
\left[
\frac{5}{16}(n\cdot m)^3-\frac{9}{16}(n\cdot m)
\right]n_\mu(1)_\nu
+
\left[
-\frac{15}{16}(n\cdot m)^2+\frac{3}{16}
\right]m_\mu(1)_\nu,
\nonumber\\
f^{(2)}_{\mu\nu,{\rm uni}}
&=&
\left[
\frac{15}{128}-\frac{45}{64}(n\cdot m)^2+\frac{35}{128}(n\cdot m)^4
\right]n_\mu(1)_\nu\nonumber\\
& &+
\left[ 
\frac{15}{32}(n\cdot m)-\frac{35}{32}(n\cdot m)^3
\right]m_\mu(1)_\nu.
\end{eqnarray}
Here we note that the terms with $n_\mu m_\nu-n_\nu m_\mu$ have vanished.
By using 
\begin{eqnarray}
& &[n_\mu (1)_\alpha-(1)_\mu n_\alpha][n^\alpha (1)_\nu-(1)^\alpha n_\nu]
=-n_\mu n_\nu+(1)_\mu (1)_\nu,
\nonumber\\
& &[m_\mu (1)_\alpha-(1)_\mu m_\alpha][m^\alpha (1)_\nu-(1)^\alpha m_\nu] 
=-m_\mu m_\nu+(1)_\mu (1)_\nu,
\nonumber\\
& &[n_\mu (1)_\alpha-(1)_\mu n_\alpha][m^\alpha (1)_\nu-(1)^\alpha m_\nu]
=-n_\mu m_\nu-(n\cdot m)(1)_\mu (1)_\nu,
\end{eqnarray}
the zero-th terms in 
$e^{-2}a^{-4}F^\mu_{\ \alpha,{\rm ret}}F^{\alpha\nu}_{\rm ret}$ 
are obtained as
\begin{eqnarray}
& &\left[
-\frac{3}{8}+\frac{9}{4}(n\cdot m)^2-(n\cdot m)^4
\right]n^\mu n^\nu 
+
\left[
\frac{3}{16}-3(n\cdot m)^2+5(n\cdot m)^4
\right](1)^\mu (1)^\nu
\nonumber\\
& &+
\left[
-\frac{9}{8}(n\cdot m)+2(n\cdot m)^3
\right](n^\mu m^\nu +m^\mu n^\nu )
+
\left[
\frac{3}{16}-\frac{3}{2}(n\cdot m)^2
\right]m^\mu m^\nu,
\label{eq:15}
\end{eqnarray}
and the zero-th terms in
$e^{-2}a^{-4}F_{\mu \nu,{\rm ret}}F^{\nu\mu}_{\rm ret}$ are obtained as
\begin{eqnarray}
\frac{3}{8}-6(n\cdot m)^2+10(n\cdot m)^4.
\label{eq:16}
\end{eqnarray}
Substituting these into the expression of the stress-energy tensor
\begin{eqnarray}
T^{\mu\nu}&=&
\frac{1}{4\pi}\left(
F^{\mu\alpha}F_\alpha^{\ \nu}
-\frac{1}{4}g^{\mu\nu}F_{\alpha\beta}F^{\beta\alpha}
\right),
\label{eleT}
\end{eqnarray}
we obtain the zero-th terms in the expansion of $T^{\mu\nu}$ 
in powers of $\epsilon$. From Eq.~(\ref{eq:15}), it is found that 
the zero-th terms of the Poynting vector $T^{0i}$, 
where the Roman index $i$ represents a spatial component, are zero. 
This result is explained by the well-known fact 
that, in the rest frame of a point charge with uniform acceleration, 
the Poynting vector of the retarded field vanishes because of the null 
magnetic field~\cite{rf:16}.

We wish to evaluate the zero-th terms in the expansion at the 
position of the charge,
but Eqs.~(\ref{eq:15}), and~(\ref{eq:16}) are indefinite 
at that position because of the angular dependence $n^\mu$. 
So let us proceed in the same manner as in the method used in 
the previous section.
That is, let us consider the expansion, in powers of $\epsilon$, 
of {\it the angular average of  $T^{\mu\nu}$ around the point charge},
and calculate the zero-th terms of the expansion. The average is taken 
in the rest frame of the charge. Using the equations 
\begin{eqnarray}
\overline{n_z^2}=\frac{1}{3},\ \ \ \ 
\overline{n_z^4}=\frac{1}{5},\ \ \ \ 
\overline{n_z^6}=\frac{1}{7},\ \ \ \ 
\overline{n_x^2 n_z^2}=\frac{1}{15},\ \ \ \ 
\overline{n_x^2 n_z^4}=\frac{1}{35}.
\label{formula}
\end{eqnarray}
for evaluating Eqs.~(\ref{eq:15}) and  
(\ref{eq:16}), and noting $(n\cdot m)=-n_z$, 
we have
\begin{eqnarray}
e^{-2}a^{-4}\overline{(F^\mu_{\ \alpha,{\rm ret}}F^{\alpha\nu}_{\rm ret})_0}
&=&
\frac{1}{560}
\left[
\begin{array}{cccc}
105&0&0&0\\
0&-2&0&0\\
0&0&-2&0\\
0&0&0&-101
\end{array}
\right],
\label{smallnon}
\end{eqnarray}
and
\begin{eqnarray}
e^{-2}a^{-4}
\overline{(F_{\mu \nu,{\rm ret}}F^{\nu\mu}_{\rm ret})_0}
&=&\frac{3}{8}\label{contrast1}.
\end{eqnarray}
Inserting these into Eq.~(\ref{eleT}), we obtain
\begin{eqnarray}
\overline{T^{\mu\nu}}&=&
\sum_{i=-4}^{\infty}\overline{(T^{\mu\nu})_i}\epsilon^i,
\nonumber\\
\overline{(T^{\mu\nu})_0}&=&
\pi\alpha\cdot\frac{1}{4480\pi^2}\frac{\hbar a^4}{c^7}
\left[
\begin{array}{cccc}
105&0&0&0\\
0&101&0&0\\
0&0&101&0\\
0&0&0&-97
\end{array}
\right].
\label{Cuniform}
\end{eqnarray}

Next, let us calculate for a point charge with circular motion. 
The world line of the charge is given by Eq.~(\ref{circ}). We have
\begin{eqnarray}
(2)^\mu&=&
-\gamma^2\Omega vl_1^\mu,
\nonumber\\
(3)^\mu&=&
-\gamma^4\Omega^2 v(l_2^\mu-v(1)^\mu),
\nonumber\\
(4)^\mu&=&
\gamma^4\Omega^3 vl_1^\mu,
\nonumber\\
(5)^\mu&=&
\gamma^6\Omega^4 v(l_2^\mu-v(1)^\mu).
\end{eqnarray}
Inserting these into Eq.~(\ref{eq:13}), and noting 
$(n\cdot l_1)=-n_x$, $(n\cdot l_2)=-n_y$, we find
\begin{eqnarray}
F_{\mu\nu,{\rm ret}}&=&
ef^{(-2)}_{\mu\nu,{\rm cir}}\epsilon^{-2}+
ef^{(-1)}_{\mu\nu,{\rm cir}}\gamma^2\Omega\epsilon^{-1}+
ef^{(0)}_{\mu\nu,{\rm cir}}\gamma^4\Omega^2+
ef^{(1)}_{\mu\nu,{\rm cir}}\gamma^6\Omega^3\epsilon+
ef^{(2)}_{\mu\nu,{\rm cir}}\gamma^8\Omega^4\epsilon^2\nonumber\\
& &-(\mu\leftrightarrow\nu)+O(\epsilon^3),\nonumber\\
f^{(-2)}_{\mu\nu,{\rm cir}}&=&
n_\mu(1)_\nu,
\nonumber\\
f^{(-1)}_{\mu\nu,{\rm cir}}&=&
\frac{1}{2}vn_xn_\mu(1)_\nu+\frac{1}{2}vl_{1\mu}(1)_\nu,
\nonumber\\
f^{(0)}_{\mu\nu,{\rm cir}}&=&
\left[
\frac{3}{8}v^2n_x^2-\frac{3}{8}v^2
\right]n_\mu(1)_\nu
+
\frac{3}{4}v^2n_xl_{1\mu}(1)_\nu
+\frac{1}{2}vn_\mu l_{2\nu}-\frac{2}{3}vl_{2\mu}(1)_\nu,
\nonumber\\
f^{(1)}_{\mu\nu,{\rm cir}}&=&
\left[
\frac{5}{16}v^3n_x^3-\frac{9}{16}v^3n_x+\frac{1}{8}vn_x
\right]n_\mu(1)_\nu
+
\left[
\frac{1}{2}v^2n_y+\frac{1}{3}v
\right]n_\mu l_{1\nu}
\nonumber\\
& &+
\left[
\frac{15}{16}v^3n_x^2-\frac{2}{3}v^2n_y-\frac{3}{16}v^3-\frac{3}{8}v
\right]l_{1\mu}(1)_\nu
\nonumber\\
& &
+
\frac{3}{4}v^2n_xn_\mu l_{2\nu}-\frac{4}{3}v^2n_xl_{2\mu}(1)_\nu
+\frac{1}{4}v^2l_{1\mu}l_{2\nu},
\nonumber\\
f^{(2)}_{\mu\nu,{\rm cir}}&=&
\left[
\frac{35}{128}v^4n_x^4-\frac{45}{64}v^4n_x^2+\frac{5}{16}v^2n_x^2
-\frac{5}{24}v^2n_y^2+\frac{15}{128}v^4+\frac{2}{5}v^3n_y
-\frac{1}{15}vn_y
\right.\nonumber\\  
& &
\left.
+\frac{5}{48}v^2
\right]
n_\mu(1)_\nu
+
\left[
\frac{5}{4}v^3n_xn_y+v^2n_x
\right]n_\mu l_{1\nu}
\nonumber\\
& &
+
\left[
\frac{35}{32}v^4n_x^3-2v^3n_xn_y-\frac{5}{4}v^2n_x-\frac{15}{32}v^4n_x
\right]l_{1\mu}(1)_\nu
\nonumber\\
& &+
\left[
\frac{15}{16}v^3n_x^2-\frac{2}{3}v^2n_y-\frac{3}{16}v^3-\frac{1}{8}v
\right]n_\mu l_{2\nu}
\nonumber\\
& & 
+
\left[
-2v^3n_x^2+\frac{5}{6}v^2n_y+\frac{1}{5}v^3+\frac{2}{15}v
\right]l_{2\mu}(1)_\nu
+\frac{5}{8}v^3n_xl_{1\mu}l_{2\nu}.
\end{eqnarray}
The terms of zero-th order in 
$e^{-2}\gamma^{-8}\Omega^{-4}F^{\mu\alpha}F_{\alpha}^{\ \nu}$ are
\begin{eqnarray}
& &\left[
-v^4n_x^4+\frac{9}{4}v^4n_x^2-\frac{3}{4}v^2n_x^2+\frac{5}{12}v^2n_y^2
-\frac{3}{8}v^4-\frac{4}{5}v^3n_y+\frac{2}{15}vn_y+\frac{1}{24}v^2
\right]n^\mu n^\nu
\nonumber\\
& &+
\left[
5v^4n_x^4-3v^4n_x^2-2v^2n_x^2+\frac{5}{4}v^2n_y^2+\frac{3}{16}v^4
+\frac{31}{30}v^3n_y+\frac{2}{15}vn_y
+\frac{5}{18}v^2
\right.\nonumber\\
& &
\hspace{0.5cm}
\left.
-\frac{21}{2}v^3n_x^2n_y
\right](1)^\mu(1)^\nu
+
\left[
v^3n_xn_y+\frac{4}{3}v^2n_x
\right]((1)^\mu l_1^\nu+l_1^\mu (1)^\nu)
\nonumber\\
& &+
\left[
-3v^3n_x^2n_y-\frac{7}{6}v^2n_x^2+\frac{2}{3}v^2n_y^2+\frac{1}{8}v^3n_y
+\frac{1}{8}vn_y+\frac{1}{6}v^2
\right](n^\mu (1)^\nu+(1)^\mu n^\nu)
\nonumber\\
& &+
\left[
-2v^4n_x^3+\frac{7}{3}v^3n_xn_y+\frac{11}{8}v^2n_x+\frac{9}{8}v^4n_x
\right](n^\mu l_{1}^\nu+l_1^\mu n^\nu)
\nonumber\\
& &+
\left[
3v^3n_x^2-v^2n_y-\frac{1}{4}v^3-\frac{1}{8}v
\right](l_2^\mu (1)^\nu+(1)^\mu l_2^\nu)
\nonumber\\
& &+
\left[
\frac{35}{12}v^3n_x^2-\frac{13}{12}v^2n_y-\frac{9}{20}v^3-\frac{2}{15}v
\right](n^\mu l_2^\nu+l_2^\mu n^\nu)
\nonumber\\
& &+
\left[
-\frac{3}{2}v^4n_x^2+\frac{2}{3}v^3n_y+\frac{3}{16}v^4+\frac{3}{8}v^2
\right]l_1^\mu l_1^\nu
+\frac{7}{6}v^3n_x(l_1^\mu l_2^\nu+l_2^\mu l_1^\nu)
-\frac{7}{36}v^2l_2^\mu l_2^\nu, 
\nonumber\\
\end{eqnarray}
and the terms of zero-th order in 
$e^{-2}\gamma^{-8}\Omega^{-4}F_{\mu\nu}F^{\nu\mu}$ are
\begin{eqnarray}
10v^4n_x^4-6v^4n_x^2-4v^2n_x^2+3v^2n_y^2
+\frac{3}{8}v^4+\frac{31}{15}v^3n_y
+\frac{4}{15}vn_y+\frac{1}{18}v^2-21v^3n_x^2n_y.\nonumber\\
\end{eqnarray}
Their angular averages are
\begin{eqnarray}
& &
\frac{1}{e^2\gamma^8\Omega^4}
\overline{(F^{\mu\alpha}F_{\alpha}^{\ \nu})_0}
=\frac{v^2}{5040}
\times\nonumber\\
& &
\left[
\begin{array}{cccc}
1085-945\gamma^{-2}&0&(2562-2982\gamma^{-2})v^{-1}&0\\
0&5055+909\gamma^{-2}&0&0\\
(2562-2982\gamma^{-2})v^{-1}&0&-4400+18\gamma^{-2}&0\\
0&0&0&-60+18\gamma^{-2}
\end{array}
\right],\nonumber\\
\end{eqnarray}
and
\begin{eqnarray}
e^{-2}\gamma^{-8}\Omega^{-4}\overline{(F_{\mu\nu}F^{\nu\mu})_0}
&=&\frac{v^2}{72}[7-27\gamma^{-2}].
\label{contrast2}
\end{eqnarray}
Substituting these into Eq.~(\ref{eleT}), we find that the zero-th term 
in the expansion of the angular averaged stress-energy tensor is
\begin{eqnarray}
& &\overline{(T^{\mu\nu})_0}
=\pi\alpha\cdot
\frac{1}{40320\pi^2}\frac{\hbar\gamma^8\Omega^4v^2}{c^5}
\times\nonumber\\
& &
\left[
\begin{array}{cccc}
1925-945\gamma^{-2}&0&(5124-5964\gamma^{-2})c/v&0\\
0&10355+873\gamma^{-2}&0&0\\
(5124-5964\gamma^{-2})c/v&0&-8555-909\gamma^{-2}&0\\
0&0&0&125-909\gamma^{-2}
\end{array}
\right].
\nonumber\\
\label{Ccirc}
\end{eqnarray}

\section{Discussion}
\label{S5}

Before comparing the self-stress with the vacuum stress, we should 
comment on the physical relevance of the self-stress which we have 
calculated. It should be noted that, although zero-th terms in the 
expansion of the self-stress themselves would be mathematically 
well defined quantities, our evaluation includes an artifical procedure 
in which we average the angular dependence of the quantities. (Although 
Teitelboim was able to rederive the radiation reaction force of 
the Lorentz-Dirac equation by applying this angular average, 
there is no guarantee
that the method of angular average is valid even in the evaluation 
of the self-stress.) 
Moreover, even if we can obtain a natural definition 
of the self-stress at the position of the particle, 
it is not clear whether we can give this quantity the definite
meaning when the physical predictability is considered.
Our interest in the calculation
is limited whether we can find out the trace 
of the Unruh-like effect in the calculation involving the self-field.

Let us compare the vacuum stress, Eq.~(\ref{Quniform}) or 
Eq.~(\ref{eq:8}), with the self-stress, Eq.~(\ref{Cuniform}) or 
Eq.~(\ref{Ccirc}). In uniform acceleration, both of 
them are proportional to the 4th power of the acceleration $a$.
In the case of circular motion, both of them are proportional to the 
4th power of $\gamma^2\Omega$, and the degrees of $v$ are equal. 
Therefore, roughly speaking, the order of vacuum stress 
multiplied by $\pi\alpha$ is equivalent to that of the self-stress. Let 
us now consider the situation in more detail. In uniform acceleration, 
while the vacuum 
stress represents an isotropic thermal bath of photons, as for the 
self-stress, the magnitude of the radiation pressure is close to 
the energy density. Moreover, 
the radiation pressure is anisotropic, 
and the tension acts along the direction of the acceleration. In circular 
motion, the self-stress represents tension along the $y$ axis.
(The signs of the components of the stress-energy tensor are
often changed if we choose another method of evaluation. 
See the following paragraph and Appendix B.) 
Furthermore, renormalization of $F_{\mu\nu}F^{\nu\mu}$ for both uniformly
accelerated and circular motion (Eqs.~(\ref{contrast1}) and 
(\ref{contrast2})) gives nonzero values, in contrast to the
fact that the vacuum expectation value of $F_{\mu\nu}F^{\nu\mu}$ 
(Eq.~(\ref{eq:4})) is zero for arbitrary motion of the observer.
Therefore, the resemblance between the two stress-energy tensors is not 
perfect. However, it would be rather impressive that the degrees of $a$, 
$\gamma^2\Omega$ and $v$, are {\it all} equal. 
One cannot discard the possibility that the self-stress which we have 
calculated reflects some indication of the Unruh-like effect.

We note that, in the derivation of the self-stress, we have used an 
expansion in powers of the distance from the charge. However, alternatively, 
we could construct the expansion in powers of the retarded time by 
substituting $z(\tau_0)$ for $x$ in Eq.~(\ref{eq:12}). 
If this is done, expansion coefficients do not include the directional 
unit vector $n^\mu$, and thus the angular dependence disappears. 
We discuss this alternative method in Appendix B.
In this method, we find, similarly, that the order of the 
vacuum stress multiplied
by $\pi\alpha$ is equivalent to that of the self-stress. 
Furthermore, we can consider the spectrum of 
the self-stress in this case, because the Fourier transform 
with respect to time can be taken,
and in fact we do so in the Appendix. 
But our trial calculation for the uniform acceleration
does not lead to a clear spectrum of the thermal bath, 
and it results in a rather awkward form.

As stated above, the physical relevance of the self-stress which 
we have evaluated is not clear. 
If one wish to investigate the more detailed relation between bremsstrahlung
and the Unruh-like effect, our approach would not be so effective,
in spite of a very long calculation.
But the resemblance between the self-stress and the vacuum stress we have 
revealed might offer a hint to investigate 
this subject.

\begin{center}
\ \\
{\Large {\bf Acknowledgements}}
\end{center}

We would like to thank S. R. Mane for helpful comments and discussions. 

\ \\

\begin{center}
\ \\
{\Large {\bf Appendix A}}\\
\vspace{0.2cm}
---{\it Expansion of a Field in Powers of $\epsilon$}---
\end{center}

In this Appendix, we perform the derivation of Eq.~(\ref{eq:13}). For 
simplicity, we set       $\tau_0=0$. First, we consider a Taylor 
expansion of the right-hand side of Eq.~(\ref{eq:12}) with respect to 
the proper time, around $\tau=\tau_0=0$. Next, we translate this 
expansion into an expansion in powers of $\epsilon$. 
In the following, the 
coefficients of the expansion are evaluated at $\tau=0$, and the time 
is omitted. 

\ \\
{\large {\bf A.1.} {\it Taylor expansion of the field}}\\

Let us expand $z(\tau)-x$ and $\dot{z}(\tau)$ around
$\tau=0$. Because the retarded time $\tau$ depends on $\epsilon$ 
in the form $\tau\sim-\epsilon$, we can write, while keeping the 
order of $\epsilon$ in mind,

\begin{eqnarray}
z(\tau)-x&=&
-\gamma+(1)\tau+\frac{1}{2}(2)\tau^2+\frac{1}{6}(3)\tau^3
+\frac{1}{24}(4)\tau^4+\frac{1}{120}(5)\tau^5
+O(\epsilon^6),\nonumber\\
\label{Gu}
\\
\dot{z}(\tau)&=&
(1)+(2)\tau+\frac{1}{2}(3)\tau^2
+\frac{1}{6}(4)\tau^3+\frac{1}{24}(5)\tau^4
+O(\epsilon^5).
\label{Tyoki}
\end{eqnarray}
Using these equations, we find
\begin{eqnarray}
\dot{z}(\tau)\cdot[z(\tau)-x]
&=&
[1-\Delta_2\epsilon]\tau-\frac{1}{2}\Delta_3\epsilon\tau^2
-\frac{1}{6}[\alpha_2+\Delta_4\epsilon]\tau^3
-\frac{1}{48}[5\dot{\alpha}_2+2\Delta_5\epsilon]\tau^4
\nonumber\\
& &+\frac{1}{240}[-9\ddot{\alpha}_2+2\alpha_3]\tau^5
+O(\epsilon^6),
\label{XYZ}
\end{eqnarray}
 where we use the following equations:
\begin{eqnarray}
(1)\cdot(1)&=&1,\nonumber\\
(1)\cdot(2)&=&0,\nonumber\\
(1)\cdot(3)&=&-\alpha_2,\nonumber\\
(2)\cdot(3)&=&\frac{1}{2}\dot{\alpha}_2,\nonumber\\
(1)\cdot(4)&=&-\frac{3}{2}\dot{\alpha}_2,\nonumber\\
(2)\cdot(4)&=&\frac{1}{2}\ddot{\alpha}_2-\alpha_3,\nonumber\\
(1)\cdot(5)&=&-2\ddot{\alpha}_2+\alpha_3.
\end{eqnarray}
We should note that, because we keep the order of $\epsilon$ in mind, 
the coefficient with  $\tau^5$ in Eq.~(\ref{XYZ}) is not equivalent 
to the fifth order term in the Taylor expansion with respect to  $\tau$.
By using the above expression, the expansion of the reciprocal of 
Eq.~(\ref{XYZ}), keeping the order of $\epsilon$ in mind, is obtained as
\begin{eqnarray}
\frac{\tau}{\dot{z}(\tau)\cdot[z(\tau)-x]}&=&
f_{(0)}+f_{(1)}\tau+\frac{1}{2}f_{(2)}\tau^2+
\frac{1}{6}f_{(3)}\tau^3+\frac{1}{24}f_{(4)}\tau^4
+O(\epsilon^5),
\nonumber\\
f_{(0)}&=&
1+\Delta_2\epsilon+(\Delta_2)^2\epsilon^2+(\Delta_2)^3\epsilon^3
+(\Delta_2)^4\epsilon^4,
\nonumber\\
f_{(1)}&=&
\frac{1}{2}\Delta_3\epsilon+\Delta_2\Delta_3\epsilon^2
+\frac{3}{2}(\Delta_2)^2\Delta_3\epsilon^3,
\nonumber\\
f_{(2)}&=&
\frac{1}{3}\alpha_2+
\left[
\frac{1}{3}\Delta_4+\frac{2}{3}\Delta_2\alpha_2
\right]\epsilon
+
\left[
(\Delta_2)^2\alpha_2+\frac{2}{3}\Delta_2\Delta_4+\frac{1}{2}(\Delta_3)^2
\right]\epsilon^2,
\nonumber\\
f_{(3)}&=&
\frac{5}{8}\dot{\alpha}_2
+
\left[
\Delta_3\alpha_2+\frac{1}{4}\Delta_5+\frac{5}{4}\Delta_2\dot{\alpha}_2
\right]\epsilon,
\nonumber\\
f_{(4)}&=&
\frac{2}{3}(\alpha_2)^2+\frac{9}{10}\ddot{\alpha}_2-\frac{1}{5}\alpha_3.
\label{Pa}
\end{eqnarray}
By substituting Eqs.~(\ref{Gu}), (\ref{Tyoki}) and~(\ref{Pa}) into the 
part of Eq.~(\ref{eq:12}) where $d/d\tau$ acts, we fix 
the part in powers of  $\tau$ up to the order $O(\epsilon^4)$.
After this, 
we carry out the differentiation.

Now, we can construct the 
expansion of Eq.~(\ref{eq:12}) in powers of $\epsilon$ by using 
the relation
\begin{eqnarray}
\tau&=&
-\epsilon-g_{(1)}\epsilon^2-\frac{1}{2}g_{(2)}\epsilon^3
-\frac{1}{6}g_{(3)}\epsilon^4-\frac{1}{24}g_{(4)}\epsilon^5
+O(\epsilon^6),
\nonumber\\
g_{(1)}&=&
\frac{1}{2}\Delta_2,
\nonumber\\
g_{(2)}&=&
\frac{3}{4}(\Delta_2)^2-\frac{1}{3}\Delta_3+\frac{1}{12}\alpha_2,
\nonumber\\
g_{(3)}&=&
\frac{15}{8}(\Delta_2)^3-2\Delta_2\Delta_3+\frac{5}{8}\Delta_2\alpha_2
+\frac{1}{4}\Delta_4-\frac{1}{8}\dot{\alpha}_2,
\nonumber\\
g_{(4)}&=&
\frac{105}{16}(\Delta_2)^4-12(\Delta_2)^2\Delta_3
+\frac{35}{8}(\Delta_2)^2\alpha_2+\frac{5}{3}(\Delta_3)^2
+\frac{7}{48}(\alpha_2)^2-\Delta_3\alpha_2,
\nonumber\\
& &
+\frac{5}{2}\Delta_2\Delta_4
-\frac{3}{2}\Delta_2\dot{\alpha}_2-\frac{1}{5}\Delta_5
+\frac{3}{20}\ddot{\alpha}_2-\frac{1}{30}\alpha_3,
\label{tau}
\end{eqnarray}
which is derived in the following subsection.
The result is obtained as Eq.~(\ref{eq:13}). 

\ \\
{\large{\bf A.2.} {\it Expansion of $\tau$ in powers of $\epsilon$}}\\
  
Let us derive Eq.~(\ref{tau}). 
The retarded time $\tau$ depends on $\epsilon$ 
according to
\begin{eqnarray}
0&=&(z(\tau)-x)\cdot(z(\tau)-x)\nonumber\\
&=&(z(\tau)-z(0)-n\epsilon)\cdot(z(\tau)-z(0)-n\epsilon).
\label{dependence}
\end{eqnarray}
For given $x^\mu$, two solutions of Eq.~(\ref{dependence}) with 
respect to $\tau$ are possible. Of course, we select the solution 
with $\tau<0$. It is found that, if one fixes $n^\mu$ in 
Eq.~(\ref{dependence}), $\tau$ is a function of $\epsilon$. 
Expanding $(z(\tau)-x)\cdot(z(\tau)-x)$ in $\tau$ by using 
Eq.~(\ref{Gu}), we get
\begin{eqnarray}
0&=&(z(\tau)-x)\cdot(z(\tau)-x)\nonumber\\
&=&
-\epsilon^2+[1-\Delta_2\epsilon]\tau^2-\frac{1}{3}\Delta_3\epsilon\tau^3
-\frac{1}{12}[\alpha_2+\Delta_4\epsilon]\tau^4
-\left[
\frac{1}{24}\dot{\alpha}_2+\frac{1}{60}\Delta_5\epsilon
\right]
\tau^5
\nonumber\\
& &+
\left[
-\frac{1}{80}\ddot{\alpha_2}+\frac{1}{360}\alpha_3
\right]
\tau^6
+O(\epsilon^7).
\label{abcd}
\end{eqnarray}
By using the relation $\tau=-\epsilon+O(\epsilon^2)$, the evaluation of  
$(z(\tau)-x)\cdot(z(\tau)-x)$ to order $O(\epsilon^4)$ gives
\begin{eqnarray}
0&=&
-\epsilon^2+[1-\Delta_2\epsilon]\tau^2+\frac{1}{3}\Delta_3\epsilon^4
-\frac{1}{12}\alpha_2\epsilon^4+O(\epsilon^5),
\end{eqnarray}
so that we have
\begin{eqnarray}
\tau^2&=&
[1-\Delta_2\epsilon]^{-1}
\left[
\epsilon^2-\frac{1}{3}\Delta_3\epsilon^4
+\frac{1}{12}\alpha_2\epsilon^4
\right]
+O(\epsilon^5),
\\
\tau&=&
[1-\Delta_2\epsilon]^{-1/2}
\left[
-\epsilon+\frac{1}{6}\Delta_3\epsilon^3
-\frac{1}{24}\alpha_2\epsilon^3
\right]
+O(\epsilon^4).
\end{eqnarray}
From these, it follows that
\begin{eqnarray}
\tau^3&=&
-\epsilon^3-\frac{3}{2}\Delta_2\epsilon^4
-\frac{15}{8}(\Delta_2)^2\epsilon^5
+\frac{1}{2}\Delta_3\epsilon^5-\frac{1}{8}\alpha_2\epsilon^5
+O(\epsilon^6),
\nonumber\\
\tau^4&=&
\epsilon^4+2\Delta_2\epsilon^5+3(\Delta_2)^2\epsilon^6
-\frac{2}{3}\Delta_3\epsilon^6+\frac{1}{6}\alpha_2\epsilon^6
+O(\epsilon^7),
\nonumber\\
\tau^5&=&
-\epsilon^5-\frac{5}{2}\Delta_2\epsilon^6
+O(\epsilon^7).
\label{stu}
\end{eqnarray}
Using the expansion (\ref{stu}), the relation 
$\tau^6=\epsilon^6+O(\epsilon^7)$, and the expansion of 
$[1-\Delta_2\epsilon]^{-1}$ with $\epsilon$, we get from  
Eq.~(\ref{abcd}) 
\begin{eqnarray}
\tau^2&=&
\epsilon^2+\Delta_2\epsilon^3
+
\left[
(\Delta_2)^2-\frac{1}{3}\Delta_3+\frac{1}{12}\alpha_2
\right]\epsilon^4
\nonumber\\
& &+
\left[
(\Delta_2)^3-\frac{5}{6}\Delta_2\Delta_3+\frac{1}{4}\Delta_2\alpha_2
+\frac{1}{12}\Delta_4-\frac{1}{24}\dot{\alpha_2}
\right]\epsilon^5
\nonumber\\
& &+
\left[
(\Delta_2)^4-\frac{35}{24}(\Delta_2)^2\Delta_3
+\frac{1}{2}(\Delta_2)^2\alpha_2
+\frac{1}{4}\Delta_2\Delta_4-\frac{7}{48}\Delta_2\dot{\alpha}_2
\right.\nonumber\\
& &\ \ \ \left.
+\frac{1}{6}(\Delta_3)^2
-\frac{7}{72}\Delta_3\alpha_2
+\frac{1}{72}(\alpha_2)^2
-\frac{1}{60}\Delta_5+\frac{1}{80}\ddot{\alpha}_2-\frac{1}{360}\alpha_3
\right]\epsilon^6
\nonumber\\
& &+O(\epsilon^7).
\end{eqnarray}
From this, we obtain Eq.~(\ref{tau}).

\begin{center}
\ \\
{\Large {\bf Appendix B}}\\
\vspace{0.2cm}
---{\it A Trial Calculation}---
\end{center}

In this appendix, We outline our trial calculation of the self-stress
in terms of the retarded proper time mentioned in section\ref{S5}. 
 The retarded time $\tau$ in Eq.~(\ref{eq:12}) depends on the 
reference point $x$ of the field.  
Now let us consider an extension 
$f_{\mu\nu}(x,\tau)$
of the function
$F_{\mu\nu,{\rm ret}}(x)$
over the time region beyond the fixed retarded time.
We could consider arbitrary extensions of $f_{\mu\nu}(x,\tau)$ 
except in the region fixed by the retarded time. 
For the moment
we choose the extended form
\begin{eqnarray}
\hspace{-1.5cm}f_{\mu\nu}(x,\tau)
&=&
\frac{e}{\dot{z}(\tau)\cdot(z(\tau)-x)}\frac{d}{d\tau}
\frac{\dot{z}(\tau)_\mu (z(\tau)_\nu-x_\nu)-
\dot{z}(\tau)_\nu (z(\tau)_\mu-x_\mu)}{\dot{z}(\tau)\cdot(z(\tau)-x)}.
\label{extension}
\end{eqnarray}

Let us evaluate the retarded field at the position of the point charge by
\begin{eqnarray}
\lim_{\tau\rightarrow\tau_0}f_{\mu\nu}(z(\tau_0),\tau).
\end{eqnarray}
Now we set $\tau_0=0$ without loss of generality.
By using Eqs.~(\ref{Gu}) and~(\ref{Tyoki}),  we obtain the expansion of 
$f_{\mu\nu}(z(0),\tau)$ in powers of $\tau$ around $\tau=0$:
\begin{eqnarray}
e^{-1}f_{\mu\nu}(z(0),\tau)
&=&
\frac{1}{2}(2)_\mu(1)_\nu\tau^{-1}
+\frac{2}{3}(3)_\mu(1)_\nu
+\frac{3}{8}(4)_\mu(1)_\nu\tau
+\frac{1}{4}(3)_\mu(2)_\nu\tau\nonumber\\
& &
+\frac{1}{3}\alpha_2(2)_\mu(1)_\nu\tau
-(\mu\leftrightarrow\nu)+O(\tau^2).
\label{timebeki}
\end{eqnarray}
(If we choose an extension other than Eq.~(\ref{extension}), 
we obtain a different form of the expansion.)
One should note that the term of order $O(\tau^{-2})$ disappears.
This equation is similar to 
Eq.~(\ref{bekibeki}), so one could interpret that the first term of 
Eq.~(\ref{timebeki}) contributes to the infinite Coulomb mass of the 
point charge. (Because $\tau<0$, the sign of this term is equal to
the sign of the first term of Eq.~(\ref{bekibeki}).)
The second term also reproduces the radiation reaction force
of the Lorentz-Dirac equation.

Let us calculate the zero-th terms of the expansion of 
$F_{\mu\alpha}F^\alpha_{\ \nu}$ in powers of $\tau$ by using 
Eq.~(\ref{timebeki}). We can choose
$f_{\mu\alpha}(z(0),\lambda\tau)
f^\alpha_{\ \nu}(z(0),-\lambda\tau)$
($\lambda$ is an arbitrary positive constant) or
$f_{\mu\alpha}(z(0),\lambda\tau)
f^\alpha_{\ \nu}(z(0),\lambda\tau)$
or any other form which is symmetric with respect to the exchange of the
indices $\mu$ and $\nu$. Although the evaluation of the zero-th terms of 
expansion does not depend on the value of $\lambda$, the spectrum  
is affected by  $\lambda$, as we see later in an explicit
calculation. 

We now choose
$f_{\mu\alpha}(z(0),\lambda\tau)
f^\alpha_{\ \nu}(z(0),-\lambda\tau)$.
The zero-th terms of the expansions are 
\begin{eqnarray}
e^{-2}(f_{\mu\alpha}(z(0),\lambda\tau) 
f^\alpha_{\ \nu}(z(0),-\lambda\tau))_0
&=&
\frac{11}{32}\dot{\alpha}_2(2)_\mu(1)_\nu
-\frac{41}{72}\alpha_2(3)_\mu(1)_\nu
+\frac{3}{16}(4)_\mu(2)_\nu
\nonumber\\
& &\hspace{-5.8cm}
+
\left[
\frac{3}{32}\ddot{\alpha_2}-\frac{59}{144}\alpha_3+\frac{1}{6}(\alpha_2)^2
\right](1)_\mu(1)_\nu
+\frac{7}{24}\alpha_2(2)_\mu(2)_\nu
-\frac{2}{9}(3)_\mu(3)_\nu
+(\mu\leftrightarrow\nu),\nonumber\\
\label{cannot}
\end{eqnarray}
and 
\begin{eqnarray}
e^{-2}(f_{\mu\nu}(z(0),\lambda\tau)
f^{\nu\mu}(z(0),-\lambda\tau))_0
&=&
\frac{3}{8}\ddot{\alpha_2}-\frac{59}{36}\alpha_3+\frac{37}{18}(\alpha_2)^2.
\label{non-zero}
\end{eqnarray}
For uniform acceleration, we have
\begin{eqnarray}
e^{-2}a^{-4}
(f_{\mu\alpha}(z(0),\lambda\tau)
f^\alpha_{\ \nu}(z(0),-\lambda\tau))_0
&=&
-\frac{5}{24}m_\mu m_\nu+\frac{5}{24}(1)_\mu(1)_\nu,
\end{eqnarray}
and
\begin{eqnarray}
e^{-2}a^{-4}
(f_{\mu\nu}(z(0),\lambda\tau)f^{\nu\mu}(z(0),-\lambda\tau))_0
&=&\frac{5}{12}.
\end{eqnarray}
Then we obtain the zero-th term of the expansion of stress-energy tensor
in powers of $\tau$:
\begin{eqnarray}
T^{\mu\nu}=\pi\alpha\cdot\frac{5}{192\pi^2}\frac{\hbar a^4}{c^7}
\left[
\begin{array}{cccc}
1&0&0&0\\
0&1&0&0\\
0&0&1&0\\
0&0&0&-1
\end{array}
\right].
\end{eqnarray}
For circular motion, we have
\begin{eqnarray}
\frac{(f_{\mu\alpha}(z(0),\lambda\tau)
f^\alpha_{\ \nu}(z(0),-\lambda\tau))_0}{
e^2\gamma^8\Omega^4 v^2}
&=&
\frac{1}{24}[-14+5\gamma^{-2}]l_{1\mu}l_{1\nu}
\nonumber\\
& &\hspace{-2cm}
+\frac{1}{72}[74-15\gamma^{-2}](1)_\mu(1)_\nu
-\frac{1}{8}v[l_{2\mu}(1)_\nu+(1)_\mu l_{2\nu}]
-\frac{4}{9}l_{2\mu}l_{2\nu},\nonumber\\
\end{eqnarray}
and
\begin{eqnarray}
\frac{(f_{\mu\nu}(z(0),\lambda\tau)
f^{\nu\mu}(z(0),-\lambda\tau))_0}{
e^2\gamma^8\Omega^4}
&=&\frac{v^4}{36}[74-15\gamma^{-2}].
\end{eqnarray}
Then, we obtain the zero-th term of the expansion of the 
stress-energy tensor in powers of $\tau$
\begin{eqnarray}
T^{\mu\nu}&=&
\pi\alpha\cdot\frac{1}{576\pi^2}\frac{\hbar\gamma^8\Omega^4v^2}{c^5}
\nonumber\\
& &\times
\left[
\begin{array}{cccc}
74-15\gamma^{-2}&0&-18v/c&0\\
0&-10+15\gamma^{-2}&0&0\\
-18v/c&0&10-15\gamma^{-2}&0\\
0&0&0&74-15\gamma^{-2}
\end{array}
\right].
\label{CirCir}
\nonumber\\
\end{eqnarray}
Thus we have obtained stress-energy tensors which are 
roughly the same order as the vacuum stresses multiplied by $\pi\alpha$.
(Although the flux in Eq.~(\ref{CirCir}) is proportional to $v^3$,
one finds that the degrees of the parameter representation of this 
flux is equal to that of vacuum stress if one takes
$v^2=1-\gamma^{-2}$ into account.)

We note that $F_{\mu\nu}F^{\nu\mu}$ of
Eq.~(\ref{non-zero}) is not zero in general motion, 
but $F_{\mu\nu}F^{\nu\mu}$ of Eq.~(\ref{eq:4}) 
is precisely zero in any motion of the observer. Furthermore, we
should note that, while
the vacuum expectation value of $F_{\mu\alpha}F^{\alpha}_{\ \nu}$ 
includes the $\eta_{\mu\nu}$ term (see Eq.~(\ref{eq:3})),  
$F_{\mu\alpha}F^{\alpha}_{\ \nu}$ calculated in this Appendix 
(see Eq.~(\ref{cannot})) does not include $\eta_{\mu\nu}$ explicitly.
Because of this fact,
$F_{1\alpha}F^{\alpha}_{\ 1}$ and $F_{2\alpha}F^{\alpha}_{\ 2}$ are 
zero in the case of uniform acceleration in the derivation in this 
Appendix, in contrast to the case of the vacuum stress. 
We note that $F_{1\alpha}F^{\alpha}_{\ 1}$  and 
$F_{2\alpha}F^{\alpha}_{\ 2}$ in the case of uniform acceleration
evaluated in section \ref{S4} (Eq.~(\ref{smallnon})) have small, 
but nonzero values which come
from the {\it angular average} of the $n^\mu n^\nu$ term in 
Eq.~(\ref{eq:15}). Therefore, the angular average method of section \ref{S4}
may induce an $\eta_{\mu\nu}$-like contribution when one evaluates
$F_{\mu\alpha}F^{\alpha}_{\ \nu}$.

Next let us calculate the spectrum of the self-stress in the case of
uniform acceleration. We have 
\begin{eqnarray}
T^{\mu\nu}&=&\lim_{\tau\rightarrow 0}
\frac{f^{03}(z(0),\lambda\tau)
f^{03}(z(0),-\lambda\tau)}{8\pi}
\left[
\begin{array}{cccc}
1&0&0&0\\
0&1&0&0\\
0&0&1&0\\
0&0&0&-1
\end{array}
\right].
\end{eqnarray}
Thus all we have to do is to calculate the Fourier transform of 
$f^{03}(z(0),\lambda\tau)f^{03}(z(0),-\lambda\tau)$.
We define 
\begin{eqnarray}
G(\tau)=e^{-2}f^{03}(z(0),\tau)f^{03}(z(0),-\tau).
\end{eqnarray}
Then it follows that
\begin{eqnarray}
\lim_{\tau\rightarrow 0}G(\lambda\tau)
&=&
\int_{-\infty}^{\infty}d\tau \delta(\tau) G(\lambda\tau)
\nonumber\\
&=&\frac{1}{2\pi}\int_{-\infty}^{\infty}d\tau 
\int_{-\infty}^{\infty}d\omega e^{i\omega\tau}
G(\lambda\tau)
\nonumber\\
&=&
\frac{1}{2\pi}\int_0^{\infty}d\omega\int_{-\infty}^{\infty}d\tau
e^{i\omega\tau}[G(\lambda\tau)+G(-\lambda\tau)].
\label{perform}
\end{eqnarray}
$G(\tau)$ has a pole at $\tau=0$. We now move this pole above the real
axis of complex $\tau$ plane. Hence $G(\tau)$ takes the form
\begin{eqnarray}
G(\tau)&=&
-\frac{a^4[1-\cosh(a\tau)]^2}{\sinh^6[a(\tau-i\epsilon)]},
\end{eqnarray}
where $\epsilon$ is an infinitesimal positive number.
This function has a periodicity $G(\tau+2\pi a^{-1} i) = G(\tau)$. 
By using this property, we can easily perform the integration over $\tau$
in Eq.~(\ref{perform})
(see section 4.4 of Ref.~\cite{rf:6}).
Although the calculation is similar to that of the Unruh effect,  
the residue of $G(\tau)$ at the pole $\tau=\pi i/ a$ causes the result
to take a rather awkward form. We obtain
\begin{eqnarray}
\lim_{\tau\rightarrow 0}G(\lambda\tau)=
\frac{1}{2\lambda^2}\int_0^{\infty}d\omega
& &\left[
\frac{\omega a^2}{2}
+\left(
\frac{11}{10}+\frac{2}{3}\frac{\omega^2}{\lambda^2 a^2}
+\frac{1}{15}\frac{\omega^4}{\lambda^4 a^4}
\right)
\frac{2\omega a^2}{e^{\pi\omega/\lambda a}-1}
\right.
\nonumber\\
& &\hspace{2cm}
\left.
-\left(
\frac{3}{5}+\frac{2}{3}\frac{\omega^2}{\lambda^2 a^2}
+\frac{1}{15}\frac{\omega^4}{\lambda^4 a^4}
\right)
\frac{2\omega a^2}{e^{2\pi\omega/\lambda a}-1}
\right].\nonumber\\
\label{awkward}
\end{eqnarray}
In this spectrum, the Planckian terms 
$\omega^3(e^{\pi\omega/\lambda a}-1)$
and $\omega^3(e^{2\pi\omega/\lambda a}-1)$, which correspond to the
temperatures $\lambda a/\pi$ and $\lambda a/2\pi$, respectively, 
appear. The first term, $\omega a^2/2$, of the spectrum resembles 
the term $\omega a^2/2$ in Eq.~(\ref{Planck}), which comes from the 
contribution of zero-point energy
$\omega(\omega^2+a^2)/2$. However, we cannot suggest 
with confidence that these results reflect some facts of real physics, 
because our evaluation here is, 
at present, rather artifical and awkward.

Finally, let us note the result of calculation in the case of
$f_{\mu\alpha}(z(0),\lambda\tau)
f^\alpha_{\ \nu}(z(0),\lambda\tau)$.
We find that for uniform acceleration,
\begin{eqnarray}
T^{\mu\nu}=\pi\alpha\cdot\frac{5}{192\pi^2}\frac{\hbar a^4}{c^7}
\left[
\begin{array}{cccc}
-1&0&0&0\\
0&-1&0&0\\
0&0&-1&0\\
0&0&0&1
\end{array}
\right],
\end{eqnarray}
and for circular motion,
\begin{eqnarray}
T^{\mu\nu}&=&
\pi\alpha\cdot\frac{1}{576\pi^2}\frac{\hbar\gamma^8\Omega^4v^2}{c^5}
\nonumber\\
& &\times
\left[
\begin{array}{cccc}
-10+15\gamma^{-2}&0&18v/c&0\\
0&74-15\gamma^{-2}&0&0\\
18v/c&0&-74+15\gamma^{-2}&0\\
0&0&0&-10+15\gamma^{-2}
\end{array}
\right].\nonumber\\
\end{eqnarray}
The spectrum of the stress-energy tensor in the case of uniform 
acceleration is just the opposite of  Eq.~(\ref{awkward}).

\end{document}